# A robust benchmark for the h- and g-indexes[1]

**Giovanni Abramo**
*National Research Council of Italy and*
*Dipartimento di Ingegneria dell'Impresa, University of Rome "Tor Vergata," Via del Politecnico 1,00133 Rome, Italy. E-mail: abramo@disp.uniroma2.it*

**Ciriaco Andrea D'Angelo and Fulvio Viel**
*Dipartimento di Ingegneria dell'Impresa, University of Rome "Tor Vergata," Via del Politecnico 1,00133 Rome, Italy.*

**Abstract**

The use Hirsch's h-index as a joint proxy of the impact and productivity of a scientist's research work continues to gain ground, accompanied by the efforts of bibliometrists to resolve some of its critical issues, through the application of a number of more or less sophisticated variants. However, the literature does not reveal any appreciable attempt to overcome the objective problems of measuring h-indexes on a large scale, for purposes of comparative evaluation. Scientists may succeed in calculating their own h-indexes but, being unable to compare them to those of their peers, they are unable to obtain truly useful indications of their individual research performance. This study proposes to overcome this gap, measuring the h- and Egghe's g-indexes of all Italian university researchers in the hard sciences, over a 5-year window. Descriptive statistics are provided concerning all of the 165 subject fields examined, offering robust benchmarks for those who wish to compare their individual performance to those of their colleagues in the same subject field.

**Introduction**

The h-index was first introduced as a "measure" of the accomplishments of an entire scientific career[2]. However, many scientists cannot resist the temptation to use it to measure their performance in research-in-progress and compare it with that of their peers. While it is true that the h-index has the attraction of providing a joint measure of the productivity and the impact of a scientist's work, and while the index is quite readily measurable for a single author, this is not at all the case if one wishes to extend the measurement to a population of scientists in a subject field, sufficiently numerous to offer a robust benchmark. Otherwise there is no explanation of why there are still no statistics for the h-index by subject field, as there are for other measures such as average citations per article, where Thomson Scientific calculates the pertinent data for each subject category. Just as for an elite runner, a scientist will not be content to know only his or her own results, because the drive to improve is partially dependant on the opportunity to compare with one's peers. Such comparisons are even more essential for the decision maker who, to allocate resources in an efficient manner, must not only compare scientists within the same subject field, but also among different subject fields. The elite runner will also want to know if peers' results have been achieved under similar conditions (wind, altitude, etc.). In terms of scientists' production, the analogous need is one of knowing if the period of comparison is the same, if the more than one scientist participated in the output (co-authorship), and if the resources used and the potential economic rents were the same. The numerous variants of the h-index (and variants of the variants) that follow one another in the literature represent attempts to perfect the indicator, rendering it as objective and representative of scientific performance as possible. Without evaluating their merits, we cite three examples: i) the individual $h_I$-index normalized by the average number of co-authors in the $h$ core, introduced by Batista et al. (2006), subsequently refined by Schreiber (2008) who introduced the fractionalized counting of the papers in his $h_m$-index, to take into account co-authorships; ii) the generalized h-index proposed by Radicchi et al. (2008), in which, to permit comparability of the index among different disciplines, the citations of each article are normalized by the average number of citations per article in the subject category of the article under observation; iii) the g-index, which, to take into account the citation evolution of the most cited papers, represents the highest number $g$ of articles that together received $g^2$ or more citations (Egghe, 2006a; Egghe, 2006b).

---

[2] In Hirsch's words (2005): "I propose the index $h$, defined as the number of papers with citation number $\geq h$, as a useful index to characterize the scientific output of a researcher.
.

Among the bibliometrists who continue to formulate ever more sophisticated indicators of performance, there are few who actually take on the difficulties of their subsequent measure. As is well known by those who devote themselves to author searches in bibliometric databases such as the Web of Science (WoS) of Thomson Reuters or Elsevier's Scopus, the problem of homonyms[3] among the scientist names makes the measure of indicators such as the h-index highly uncertain and unreliable, unless it is the author oneself who is calculating his or her own index. For this reason, Bornmann and Daniel (2007) recommend "calculating the h-index on the basis of a complete list of publications that is authorized by the scientist himself or herself". However, a full-scale application of this recommendation would be difficult to accomplish, because the manual extraction of the publications identified by the individual authors would be extremely time-consuming, as would be the process of each author checking that others had made the correct selection of his or her own publications, assuming that all the authors would be available to take on the task. In addition, one would have to depend on the accuracy and honesty of individual scientists in not listing publications that could be erroneously attributed to them. When the audit company KPMG conducted audits of publication lists submitted by universities, for the 1997 Australian research assessment exercise, the examination found a high error rate (34%). KPMG also found that 97% of the errors affected final scores, and therefore also funding allocations (Harman, 2000).

---

[3] The authors estimated 12% of homonyms among Italian university scientists (Abramo et al., 2010).

Some h-index and *h* variant statistics found in the literature are a byproduct of analyses of convergent validity. These are generally restricted to a limited number of observations and scientific disciplines. There are, for example, Bormann et al. (2008), who use data concerning 414 postdoctoral researchers in biomedicine (publication window 1996-2004); van Raan (2006) measuring the h-index of 147 university research groups, comprising 700 researchers in chemistry and chemical engineering in the Netherlands (publication window 1991-2000); Kelly and Jennions (2006) measuring the h-index of 187 individual editorial board members (ecologists and evolutionary biologists) of seven journals; Cronin and Meho (2006), ranking 31 influential information science scientists, mainly in their mid- to late-career. Other elaborations concern h-index rankings of living scientists, which are found to favor older scientists. Schaefer and Peterson (2007) elaborated a list of living chemists with h-index above 49.

A similar one is available for economists registered with RePEc (research papers in economics)[4].

Our study sets out to measure the h-index and *g* index for all Italian university researchers in the hard sciences, grouped according to the subject fields identified for them under the formal Italian university classification system. The publications examined are extracted from the WoS, and the window of publication is 5 years: 2001-2005. The date for observing the number of citations received by the publications is March 31, 2008. Although the number of journals documented by Thomson varies from year to year, meaning that the probability of citations being recorded is also variable, the statistics furnished are sufficient to constitute a robust benchmark for all those scientists who wish a comparison for their own performance over a 5-year period, after the lapse of 27 months from the close of the period.

**Dataset and methodology**

Data used in this study are taken from the Observatory on Public Research in Italy (ORP), a bibliometric database derived by the authors from the WoS. The ORP provides a census of scientific production from all research institutions situated in Italy. Beginning from this database, the next step was to extract the publications authored by the scientists of all 79 Italian universities, in the period 2001-2005, which is a total of roughly 147,000 works. Through the development of a complex algorithm for recognition and reconciliation of addresses and the disambiguation of the real identity of the authors, it was then possible to accurately attribute each publication to the university researchers who wrote it. This step is necessary because, with bibliometric databases such as Elsevier's Scopus and Thomson Reuters' Web of Science, it is actually a formidable task to identify the true authors of publications. For the period under observation, the "authors' list" and the "address list" are not linked in such a manner as to identify the home institution of the authors. In addition, only the author's last name and first name initials are reported. As a consequence, any time the address list indicates two or more institutions, one does not know immediately to which one each author belongs. In addition, the rate of homonyms is very high among large populations of scientists: 12% of the 60,000 scientists in the Italian university system have last names that are homonyms of those for other scientists. Eliminating ambiguities as to the precise identity of the author within acceptable margins of error[5] is a daunting task, explaining why bibliometric studies are generally carried out at aggregated level of analysis, such as at the university level. When the analyses presented in the literature are based on examinations at the level of single scientists or research groups they are generally limited to a maximum of a few organizations or scientific disciplines. In these cases it is possible to disambiguate manually.

---

[4] http://ideas.repec.org/top/top.person.hindex.html

For greater significance, the current analysis proceeded by dividing the field of observation according to the formal "scientific disciplinary sectors", or SDSs of the Italian university system, in which each scientist belongs. The study refers to all hard science SDSs (165 in total) where at least 50% of the scientists had published at least one scientific article in the period under examination. These SDSs are in turn grouped into nine formal "university disciplinary areas", or "UDAs" (details of the classification system are provided on http://www.disp.uniroma2.it/laboratorioRTT/TESTI/Indicators/ssd.html). In the period under examination there were 37,606 scientists on staff in the 165 SDSs considered. The scientists were identified from the CINECA database of the Italian Ministry of Universities and Research (http://cercauniversita.cineca.it/php5/docenti/cerca.php). The study excludes those scientists who entered the university system after 2001 or departed prior to 2005. The dataset is therefore composed of the 27,502 researchers who remained in role, throughout the five-year period. For those who changed SDSs during this period, provided they remained in the hard sciences, the scientist was "assigned" to the SDS to which he or she belonged at the close of 2005.

---

[5] The harmonic mean of precision and recall (F-measure) for identifying authorship, as disambiguated by our algorithm, is around 95% (2% margin of error, 98% confidence interval).

**Results**

The 6,064 scientist found to have a null h-index (or *g* index) were excluded from further analysis. Of these, 4,970 had not produced any publications in the period under observation and 1,094 were authors of publications that had not received any citations.

The distribution of the indicators is summarized by statistics for the quartiles, average and variance. Table 1 presents the statistics concerning the h-index, for the nine UDAs examined.

The data show two substantially different groups of UDAs. On one hand there are Chemistry, Biology, Medicine and Physics, which are characterized by a high variance in the data, never less than 12. On the other are the remaining five UDAs (Mathematics and computer sciences, Earth sciences, Agricultural and veterinary sciences, Civil engineering and architecture, Industrial and information engineering), which have a very low variance, consistently below 5. In these latter UDAs, the h-indexes of the researchers are thus quite uniform, with very limited ranges. The value for the first quartile is always 1, while the median is always 2. The average h-index never exceeds a value of 3. In contrast, for Physics, Chemistry, Biology and Medicine, the first quartile is never less than 2 and the median is never less than 4. The average h-index oscillates between 4.94 for Medicine and 6.25 for Chemistry, and the maximum between 25 for Physics and 36 for Chemistry. The Chemistry UDA is clearly the one with the greatest dispersion of data.

Table 2 presents the data from calculations conducted for the *g* index. The statistics for the distribution of this indicator confirm what emerges from analysis of the h-index: Mathematics and computer sciences, Earth sciences, Agricultural and veterinary sciences, Civil engineering and architecture, Industrial and information engineering show significant uniformity. The value of the first quartile oscillates between 1 and 2 and the mean is always 3. For these same UDAs, the average *g* index oscillates between 3.38 for Mathematics and computer sciences and 4.52 for Agricultural and veterinary sciences. The situation is much different for the other UDAs, where the value for the first quartile is never less than 3 and the median never less than 6. The average oscillates between 7.37 for Biology and 9.18 for Chemistry, and the maximums between 43 for Physics and 58 for Biology and Medicine.

[Table 1]

[Table 2]

At a greater level of detail, the analysis by single SDSs also reveals significant unevenness within the UDAs. As an example, Table 3 presents statistics on the distribution of the h-index for the Physics UDA. There are only two SDSs where the first quartile is other than 2: FIS/03 (Physics of matter) and FIS/05 (Astronomy and astrophysics). These are also the two SDSs that register the maximum variance in performance and the highest average values: 6.29 for Physics of matter and 6.91 per Astronomy and astrophysics. The most homogenous SDS seems to be FIS/06 (Earth physics and atmospheric environment): here the third quartile shows a value of 4 and the maximum is 10. This is also the smallest SDS, with only 42 scientists having a non-null h-index. The absolute maximum is seen in FIS/03, where the top scientist registers an h-index of 25, followed by FIS/05, with a maximum at 23, and FIS/01 at 22.

Table 4 presents the statistics for the distribution of the $g$ index for the SDSs of a single UDA, again for the example of Physics. The data show a characterization of the single SDSs that is not dissimilar to that emerging from the h-index: once again, FIS/03 and FIS/05 show greater heterogeneity than the other five SDSs

The data concerning all 165 SDSs from all UDAs examined are available at:

http://www.disp.uniroma2.it/laboratoriortt/TESTI/Indicators/metodology.html

[Table 3]

[Table 4]

The above statistics demonstrate significant heterogeneity in the distribution of the h-index values, and also of those for the $g$ index, both for scientists of the various UDAs and scientists of different SDSs within the same UDA. We wish to further investigate this aspect of heterogeneity, which is particularly relevant if such indicators are used for comparative evaluation of non-homogenous groupings, precisely as in the situation of comparing researchers from different UDAs or from different SDSs of the same UDA. Table 5 presents the ranges for descriptive statistics (median and maximum) of the distributions for h-index recorded for all the researchers from all SDSs in each UDA. In Mathematics and computer sciences, the median oscillates between 1 and 2, while the maximum value is between 6 and 13. The SDSs of Civil engineering and architecture show an analogous trend. In contrast, for Biology and Medicine the medians show ranges of 5, oscillating between 1 and 6 for the Biology SDSs and from 2 to 7 for the Medicine SDSs. These UDAs also show the widest variation in maximums: 27 for medicine and 26 for Biology.

[Table 5]

Going on to the statistics for the individual SDSs of each UDA (Table 6) it emerges that in all nine of the SDSs of Mathematics and computer sciences, the first quartile is always 1 and the median never exceeds 2. Once again, the situation for Civil engineering and architecture is similar. In Physics, this flattening towards lower levels is not seen for any SDSs, while only one of the 11 Chemistry SDSs has a first quartile of 1. In Biology there are two of 19 SDSs in which the first quartile is 1, while there is only one case where the median is less than or equal to 2.

[Table 6]

The same analysis, repeated for the *g* index, does not offer substantial differences, but confirms the significant differences among UDAs in terms of heterogeneity of performance (Table 7). If anything, the variability among the SDSs increases, even for those UDAs that resulted as particularly homogenous in the h-index analysis. In Mathematics and computer sciences, the median *g* index for its nine SDSs oscillates between 2 and 3, while the maximum varies between 10 and 47. In Physics, Chemistry, Biology and Medicine we once again see the extreme heterogeneity among the SDSs, with ranges for the median of 5, 4, 6 and 9, respectively. In the Physics UDA, in particular, none of the seven SDSs register a first quartile of 1 or a median of less than 3 (Table 8).

[Table 7]

[Table 8]

**Discussion and conclusions**

In this work we have presented some statistics descriptive of the distribution of values of h-index and *g* index registered for all scientists in the hard science areas of the Italian university system, over a 5-year window. This data provides useful information for those who wish to compare their own performance with that of colleagues in the same subject field, provided they are suitably aware of the methodological assumptions involved in the dataset and the elaborations developed. The results of the analysis show significant variability in the characteristics of the distributions for the various scientific sectors, reflecting the well known fact that different sectors show different intensity of both publication and citation.

The literature has previously raised the issue of varying intensity of citations, with a number of studies examining and resolving it, especially through normalizations to the sectorial averages. To reduce the distortion provoked by varying intensity of publication, which would favor scientists employed in the more "fertile" sectors when comparisons are conducted at the individual level, studies have proceeded by limiting their analysis to homogenous groups of scientists. The current study has instead examined an entire heterogeneous national population, but with consideration of the official scientific disciplinary sector to which each researcher belongs. But it is also likely that the scientists in any given SDS sometimes carry out research in other subjects, and thus also publish in those fields. For example "statisticians" could both work in the subject field of "statistics and probability", and also in fields where this discipline is used as a tool to meet objectives of other disciplines (computational biology, epidemiology, applied physics, etc.), which could be fields offering significantly higher intensity of publication. Jorge E. Hirsch himself, between 2003 and 2008, was the author of 18 articles in WoS journals, of which nine are in *physics condensed matter*, two in *physics applied*, five in *physics multidisciplinary* and two in *information science & library science*. Does it make sense to compare his h-index with those of colleagues in his dominant sector (*physics condensed matter*), when the major determinant of his bibliometric performance has been his article on the h-index, classified as *information science & library science*?

There is an evident need for deeper investigation that analyzes the critical issue of varying sectorial intensity, in both publication and citations, and proposes a new indicator that reduces its distorting effects. On one hand, the needs for administrative efficiency may explain the adoption of simple evaluation systems and indicators (such as the h-index), but on the other hand these can have unacceptable collateral effects. It is those who are directly concerned, namely the researchers, who will demand that any systems of evaluation for the productivity and quality of their scientific work, regardless of simplicity, also be transparent, exhaustive and trustworthy.

| UDA | N. of SDS | N. of scientists | h index quartiles | | | | Average | Variance |
|---|---|---|---|---|---|---|---|---|
| | | | 1° | Median | 3° | Max | | |
| Mathematics and computer sciences | 9 | 1,732 | 1 | 2 | 3 | 13 | 2.31 | 2.23 |
| Physics | 7 | 1,846 | 2 | 4 | 7 | 25 | 5.04 | 12.50 |
| Chemistry | 11 | 2,597 | 4 | 6 | 8 | 36 | 6.24 | 13.74 |
| Earth sciences | 12 | 794 | 1 | 2 | 4 | 11 | 2.90 | 4.10 |
| Biology | 19 | 3,621 | 3 | 4 | 7 | 33 | 5.02 | 12.64 |
| Medicine | 41 | 6,277 | 2 | 4 | 7 | 33 | 4.94 | 15.56 |
| Agricultural and veterinary sciences | 25 | 1,514 | 1 | 2 | 4 | 18 | 3.03 | 4.88 |
| Civil engineering and architecture | 5 | 484 | 1 | 2 | 3 | 12 | 2.41 | 2.89 |
| Industrial and information engineering | 36 | 2,573 | 1 | 2 | 4 | 19 | 2.89 | 4.63 |

*Table 1: h index quartiles for Italian university scientists grouped by UDA*

| UDA | g index quartiles | | | | Average | Variance |
|---|---|---|---|---|---|---|
| | 1° | Median | 3° | Max | | |
| Mathematics and computer sciences | 1 | 3 | 4 | 47 | 3.38 | 9.56 |
| Physics | 3 | 6 | 11 | 43 | 7.75 | 35.45 |
| Chemistry | 5 | 8 | 12 | 55 | 9.18 | 34.39 |
| Earth sciences | 2 | 3 | 6 | 18 | 4.12 | 10.63 |
| Biology | 3 | 6 | 10 | 58 | 7.37 | 32.40 |
| Medicine | 3 | 6 | 11 | 58 | 7.62 | 46.63 |
| Agricultural and veterinary sciences | 2 | 3 | 6 | 35 | 4.52 | 13.44 |
| Civil engineering and architecture | 1 | 3 | 5 | 20 | 3.45 | 7.50 |
| Industrial and information engineering | 2 | 3 | 6 | 32 | 4.24 | 12.83 |

*Table 2: g index quartiles for Italian university scientists grouped by UDA*

| SDS | N. of scientists | h index quartiles | | | | Average | Variance |
|---|---|---|---|---|---|---|---|
| | | 1° | Median | 3° | Max | | |
| FIS/01 | 745 | 2 | 4 | 6 | 22 | 4.50 | 10.41 |
| FIS/02 | 264 | 2 | 5 | 7 | 17 | 5.14 | 11.06 |
| FIS/03 | 331 | 4 | 6 | 8 | 25 | 6.29 | 14.00 |
| FIS/04 | 133 | 2 | 4 | 6 | 11 | 4.32 | 7.57 |
| FIS/05 | 134 | 3 | 5 | 10 | 23 | 6.91 | 28.59 |
| FIS/06 | 42 | 2 | 3 | 4 | 10 | 3.21 | 4.12 |
| FIS/07 | 197 | 2 | 4 | 6 | 13 | 4.45 | 6.83 |

*Table 3: h index quartiles for Italian university scientists in the Physics UDA*

| SSD | N. of scientists | g index quartiles | | | | Average | Variance |
|---|---|---|---|---|---|---|---|
| | | 1° | Median | 3° | Max | | |
| FIS/01 | 745 | 3 | 6 | 10 | 37 | 6.99 | 28.47 |
| FIS/02 | 264 | 3 | 7 | 11 | 30 | 7.78 | 33.74 |
| FIS/03 | 331 | 5 | 9 | 12 | 43 | 9.79 | 44.20 |
| FIS/04 | 133 | 3 | 6 | 11 | 22 | 6.83 | 25.52 |
| FIS/05 | 134 | 4 | 8 | 16 | 36 | 10.52 | 73.18 |
| FIS/06 | 42 | 2 | 4 | 6 | 17 | 4.64 | 14.09 |
| FIS/07 | 197 | 3 | 6 | 9 | 20 | 6.55 | 17.66 |

*Table 4: g index quartiles for Italian university scientists in the Physics UDA*

|  | Median | | Max | |
| --- | --- | --- | --- | --- |
| UDA | Min | Max | Min | Max |
| Mathematics and computer sciences | 1 | 2 | 6 | 13 |
| Physics | 3 | 6 | 10 | 25 |
| Chemistry | 3 | 6 | 7 | 36 |
| Earth sciences | 1 | 4 | 6 | 11 |
| Biology | 1 | 6 | 7 | 33 |
| Medicine | 2 | 7 | 6 | 33 |
| Agricultural and veterinary sciences | 1 | 5 | 6 | 18 |
| Civil engineering and architecture | 2 | 3 | 6 | 12 |
| Industrial and information engineering | 1 | 5 | 4 | 19 |

*Table 5: Ranges of medians and maximums for the distribution of h indexes among the SDSs of each UDA*

| UDA | Total N. of SDSs | N. of these with first quartile = 1 | N. of these with median <= 2 |
| --- | --- | --- | --- |
| Mathematics and computer sciences | 9 | 9 | 9 |
| Physics | 7 | 0 | 0 |
| Chemistry | 11 | 1 | 0 |
| Earth sciences | 12 | 7 | 6 |
| Biology | 19 | 2 | 1 |
| Medicine | 41 | 13 | 10 |
| Agricultural and veterinary sciences | 25 | 14 | 15 |
| Civil engineering and architecture | 5 | 4 | 4 |
| Industrial and information engineering | 36 | 24 | 22 |

*Table 6: Number of SDSs where the first quartile of h index equals 1 and the median is less than or equal to 2, for each UDA*

|  | Median | | Max | |
| --- | --- | --- | --- | --- |
| UDA | Min | Max | Min | Max |
| Mathematics and computer sciences | 2 | 3 | 10 | 47 |
| Physics | 4 | 9 | 17 | 43 |
| Chemistry | 5 | 9 | 10 | 55 |
| Earth sciences | 2 | 5 | 8 | 18 |
| Biology | 2 | 8 | 10 | 58 |
| Medicine | 2 | 11 | 9 | 58 |
| Agricultural and veterinary sciences | 2 | 7 | 9 | 35 |
| Civil engineering and architecture | 2 | 4 | 11 | 20 |
| Industrial and information engineering | 1 | 7 | 7 | 32 |

*Table 7: Range of medians and maximums for the distribution of g indexes among the SDSs of each UDA*

| UDA | Total N. of SDSs | N. of these with first quartile = 1 | N. of these with median <= 2 |
| --- | --- | --- | --- |
| Mathematics and computer sciences | 9 | 5 | 4 |
| Physics | 7 | 0 | 0 |
| Chemistry | 11 | 1 | 0 |
| Earth sciences | 12 | 5 | 3 |
| Biology | 19 | 2 | 1 |
| Medicine | 41 | 6 | 1 |
| Agricultural and veterinary sciences | 25 | 9 | 5 |
| Civil engineering and architecture | 5 | 2 | 2 |
| Industrial and information engineering | 36 | 13 | 12 |

*Table 8: Number of SDSs where the first quartile of g index equals 1 and the median is less than or equal to 2, for each UDA*